\documentclass[12pt,preprintnumbers,eqsecnum]{revtex4}
\usepackage{amsfonts}
\usepackage{mathrsfs}
\usepackage{amssymb}
\usepackage{amsmath}
\usepackage{graphicx}
\usepackage{epstopdf}
\usepackage{pstricks}
\usepackage{slashed}
\usepackage{mathbbol}
\usepackage{color}

\def\be{\begin{equation}}
\def\ee{\end{equation}}
\def\bea{\begin{eqnarray}}
\def\eea{\end{eqnarray}}
\def\bean{\begin{eqnarray*}}
\def\eean{\end{eqnarray*}}
\def\bary{\begin{array}}
\def\eary{\end{array}}
\def\bit{\begin{itemize}}
\def\eit{\end{itemize}}

\def\su5u1{SU(5) \times U(1)}
\def\fsu5u1{SU(5) \times U(1)'}
\def\so10{SO(10)}
\def\sq20{SO(10) \times SO(10)}

\def\bwt{\begin{widetext}}
\def\ewt{\end{widetext}}
\def\be{\begin{equation}}
\def\ee{\end{equation}}
\def\bea{\begin{eqnarray}}
\def\eea{\end{eqnarray}}
\def\bean{\begin{eqnarray*}}
\def\eean{\end{eqnarray*}}
\def\bary{\begin{array}}
\def\eary{\end{array}}
\def\bit{\begin{itemize}}
\def\eit{\end{itemize}}

\def\su5u1{SU(5) \times U(1)}
\def\fsu5u1{SU(5) \times U(1)'}
\def\so10{SO(10)}
\def\sq20{SO(10) \times SO(10)}

\usepackage{amsmath}

\usepackage{amssymb}
\usepackage{gensymb}

\begin{document}

\setlength{\parskip}{0.01cm}

\preprint{ OSU-HEP-14-03}

\title{\Large A predictive model of Dirac Neutrinos}
\author{Shreyashi Chakdar\footnote{Electronic address: chakdar@okstate.edu},
Kirtiman Ghosh\footnote{Electronic address: kirti.gh@gmail.com} and S. Nandi\footnote{Electronic address: s.nandi@okstate.edu}}
\affiliation{Department of Physics and Oklahoma Center for High Energy Physics, 
Oklahoma State University, Stillwater OK 74078-3072, USA. \\}
%


\begin{abstract} 
\section*{Abstract}
Assuming lepton number conservation, hermiticity of the neutrino mass matrix and $\nu_{\mu} - \nu_{\tau}$ exchange symmetry, we show that we can determine the neutrino mass matrix completely
from the existing data. Comparing with the existing data, our model predicts an inverted mass hierarchy (close to a degenerate pattern) with the three neutrino mass values, $8.91 \times 10^{-2}$ eV, $8.95 \times 10^{-2}$ eV, $7.50 \times 10^ {-2}$ eV, a large value for the CP violating phase , $\delta = 110^0$, and of course, the absence of neutrinoless $\beta \beta$ decay. All of these predictions can be tested in the forthcoming or future  precision neutrino experiments.

\end{abstract}

\pacs{11.10.Kk, 11.25.Mj, 11.25.-w, 12.60.Jv}

\keywords{Dirac Neutrino; Inverse hierarchy, $\mu-\tau$ symmetry}

\maketitle

\section{Introduction}
In the past 20 years, there has been a great deal of progress  in neutrino physics from the atmospheric neutrino experiments (Super-K \cite{Wendell:2013kxa}, K2K \cite{Mariani:2008zz}, MINOS \cite{Barr:2013wta}), solar neutrino experiments ( SNO \cite{Aharmim:2011vm}, Super-K \cite{Koshio:2013dta} , KamLAND \cite{Mitsui:2011zz}) as well as  reactor/accelerator neutrino experments (Daya Bay \cite{An:2012eh}, RENO \cite{Ahn:2012nd}, Double Chooz \cite{Abe:2011fz}, T2K\cite{Abe:2011sj}, NO$\nu$a \cite{Patterson:2012zs}). These experiments have pinned down three mixing angles - $\theta_{12}$, $\theta_{23}$, $\theta_{13}$ and two mass squared differences $\Delta m^2_{ij} = m^2_{i}-m^2_{j}$ with reasonable accuracy
\cite{deGouvea:2013onf}. However there are several important  parameters yet to be measured. These include the value of the CP phase $\delta$ which will determine the magnitude of CP violation in the leptonic sector and the sign of $\Delta m^2_{32}$ which will determine whether the neutrino mass hierarchy is normal or inverted. We also don't know yet if the neutrinos are Majorana or Dirac particles.

On the theory side, the most popular mechanism for neutrino mass generation is the see-saw \cite{Gell-Mann}. This requires heavy right handed neutrinos, and this comes naturally in the $SO(10)$ grand unified theory
(GUT) \cite{Georgi} in the $16$ dimensional fermion representation. The tiny neutrino masses require the scale of these right handed neutrinos close the GUT scale.  The light neutrinos generated via the sea-saw mechanism are Majorana particles. However, the neutrinos can also be Dirac particles just like ordinary quarks and lepton. This can be achieved by adding right handed neutrinos to the Standard Model. The neutrinos can get tiny Dirac masses via the usual Yukawa couplings with the SM Higgs. In this case, we have to assume that the corresponding Yukawa couplings are very tiny, $\sim 10^{-12}$. 
Alternatively, we can introduce a 2nd Higgs doublet and a discrete  $Z_2$ symmetry so that the  neutrino masses are generated only from the 2nd Higgs doublet. 
The neutrino masses are generated from the spontaneous breaking of this discrete symmetry from a tiny vev of this 2nd Higgs doublet in the eV or keV range, and then the associated Yukawa couplings need not be so tiny \cite{Gabriel:2006ns}. At this stage of neutrino physics, we can not  determine which of these two possibilities are realized by nature.

In this work, we show that  with the three known mixing angles and two known mass difference squares, we find an interesting pattern in the neutrino mass matrix if the neutrinos are Dirac particles. With three reasonable assumptions : (i) lepton number conservation, (ii) hermiticity of the neutrino mass matrix, and (iii) $\nu_{\mu}$ - $\nu_{\tau}$ exchange symmetry, we can construct the neutrino mass matrix completely. The resulting mass matrix satisfies all the constraints implied by the above three assumptions, and gives an inverted hierarchy (IH) (very close to the degenerate) pattern. We can now predict the absolute values of the masses of the three neutrinos, as well as the value of the CP violating phase $\delta$. We also predict the absence of neutrinoless double $\beta \beta$  decay.     

\section{The Model and the neutrino mass matrix}
Our model is based on the Standard Model (SM) Gauge symmetry, $SU(3)_C \times SU(2)_L \times U(1)_Y$, supplemented by a discrete $ Z_2$ symmetry. \cite{Gabriel:2006ns}.  In addition to the SM  particles,  we have three SM singlet right handed neutrinos, $N_{Ri}$, i = 1,2,3, one for each family of fermions.  We also have one additional  Higgs doublet $\phi$, in addition to the usual SM Higgs doublet $\chi$.
 All the SM particles are even under $Z_2$, while the $N_{Ri}$ and  the  $\phi$ are odd under $Z_2$. Thus while the SM quarks and leptons obtain their masses from the usual Yukawa couplings with $\chi$ with vev of $\sim 250$ GeV, the neutrinos get masses only from its Yukawa coupling with $\phi$ for which we assume the vev is $\sim$ keV to satisfy the cosmological constraints which we will discuss later briefly. Note that even with as large as a keV vev for $\phi$, the corresponding Yukawa coupling is of order $10 ^{-4}$ which is not too different from the light quarks and leptons Yukawa coupling in the SM.
The Yukawa interactions of the Higgs fields $\chi$ and $\phi$ and the leptons can be written as,
\begin{equation}
L_Y = y_l \bar{\Psi}_L^l l_R \chi + y_{\nu l} \bar{\Psi}_L^l N_R \tilde{\Phi} + h.c.,
\end{equation}
where $\bar{\Psi}_L^l = (\bar{\nu_l}, \bar{l})_L$ is the usual lepton doublet and $l_R$ is the charged lepton singlet, and we have omitted the family indices. The
first term gives rise to the masses of the charged leptons, while the second term gives tiny 
neutrino masses. The interactions with the quarks are the same as in the Standard Model
with $\chi$ playing the role of the SM Higgs doublet. 
Note that in our model, the tiny neutrino masses are generated from the spontaneous breaking of the discrete $Z_2$ symmetry with its tiny vev of 
$\sim$ keV. The left handed doublet neutrino combine with its corresponding 
right handed singlet neutrino to produce a massive Dirac neutrino.
Since we assume lepton number conservation, the Majorana mass terms for the right handed neutrinos, having the form, $M \nu_R^TC^{-1}\nu_R$ are not allowed. 

The model has a very light neutral scalar $\sigma$ with mass of the order of this $Z_2$ symmetry breaking scale. 
Detailed phenomenology of this light scalar $\sigma$
 in context of $e+e-$ collider has been done previously \cite {Gabriel:2006ns} and also some phenomenological works have been done on the chromophobic charged Higgs of this model at the LHC whose signal are very different from the charged Higgs in the usual two Higgs doublet model \cite{Gabriel:2008es}. There are bounds on $v_{\phi}$ from cosmology, big bang nucleosynthesis, because of the presence of extra degree of freedom compared to the SM; puts a lower limit on $v_{\phi}\geq 2$ eV \cite{logan}, while the bound from supernova neutrino observation is $v_{\phi}\geq 1$ keV \cite{Raffelt:2011nc}.
 
 In this paper, we study the neutrino sector of the model using the input of all the experimental information regarding the neutrino mass difference squares and the three mixing angles. Our additional theoretical inputs are  that the neutrino mass matrix is hermitian and also has  $\nu_{\mu} - \nu_{\tau}$ exchange symmetry. We find that in order for our model to be consistent with the current available experimental data, the neutrino mass hierarchy has to be inverted type (with neutrino mass values close to degenerate case). We also predict the values of all three neutrino masses, as well as the CP violating phase $\delta$.
 
With the three assumptions stated in the introduction, namely, lepton number conservation, Hermiticity of the neutrino mass matrix, and the $\nu_{\mu} - \nu_{\tau}$ exchange symmetry, the neutrino mass matrix can be written as

\begin{equation} 
 M_{\nu} = \left(\begin{matrix}
a & b & b \\
b^* & c & d\\
b^* & d & c
\end{matrix}
\right).  
\label{eqn1}
\end{equation}
The parameters a, c and d are real, while the parameter b is complex.
Thus the model has a total of five real parameters.
The important question at this point is whether the experimental data is consistent with this form. Choosing a basis in which the Yukawa couplings for the charged leptons are diagonal, the PMNS matrix in our model is simply given by $U_{\nu}$, where $U_{\nu}$ is the matrix which diagonalizes the neutrino mass matrix. Since the neutrino mass matrix is hermitian, it can then be obtained from
\begin{equation} 
M_{\nu} = U_{\nu} \bold {M}_{\nu}^{diag} U_{\nu}^{\dagger}
\label{eqn0}
\end{equation}

where 

\begin{equation} 
 \bold {M}_{\nu}^{diag}= \left(\begin{matrix}
m_1 & 0 & 0 \\
0 & m_2 & 0\\
0 & 0 & m_3
\end{matrix}
\right). 
\end{equation}

The matrix $U_{\nu}$ is the PMNS matrix for our model (since $U_l$ is the identity matrix from our choice of basis), and  is conventionally written as: 
\begin{equation} 
U_{\nu}
= \left(\begin{matrix}
c_{12}c_{13} & s_{12}c_{13} & s_{13}e^{-i\delta} \\
-s_{12}c_{23}-c_{12}s_{23}s_{13}e^{i\delta} & c_{12}c_{23}-s_{12}s_{23}s_{13}e^{i\delta} & s_{23}c_{13} \\
s_{12}s_{23}-c_{12}c_{23}s_{13}e^{i\delta} & -c_{12}s_{23}-s_{12}c_{23}s_{13}e^{i\delta} & c_{23}c_{13}
\end{matrix}
\right),  
\label{eqn2}
\end{equation}
where, $c_{ij}={\rm Cos}\theta_{ij}$ and $s_{ij}={\rm Sin}\theta_{ij}$.

\section{Results}
The values of three mixing angles and the two neutrino mass squared differences are now determined from the various solar, reactor and accelerator neutrino experiments with reasonable accuracy (the sign of $\Delta m^2_{32}$ is still unknown).
The current knowledge of the mixing angles and mass squared differences are given by~\cite{PDG} Table \ref{tab:expdata}.
\begin{table}[htb]
\begin{center}
\begin{tabular}{|c|c|}
\hline
\hline
Parameter & best-fit ($\pm\sigma$) \\
\hline
$\Delta m^2_{21}[10^{-5} eV^2]$ & $7.53_{-0.22}^{+0.26}$ \\
\hline
$\Delta m^2[10^{-3} eV^2]$ & $2.43_{-0.10}^{+0.06}$ \\
\hline
$\sin^2\theta_{12}$ & $0.307_{-0.016}^{+0.018}$\\
\hline
$\sin^2\theta_{23}$ & $0.386_{-0.021}^{+0.024}$\\
\hline
$\sin^2\theta_{13}$ & $0.0241\pm 0.0025$\\
\hline
\end {tabular}
\end{center}
\caption {The best-fit values and $1\sigma$ allowed ranges of the 3-neutrino oscillation parameters. The definition of $\Delta m^2$ used is $\Delta m^2 = m_{3}^2 - (m_{2}^2 + m_{1}^2)/2$. Thus $\Delta m^2 = \Delta m_{31}^2 - m_{21}^2/2$ if $m_1 < m_2 < m_3$ and $\Delta m^2 = \Delta m_{32}^2 + m_{21}^2/2$ for $m_3 < m_1 < m_2$.}
\label{tab:expdata}
\end{table}
\\
It is not at all sure that the data will satisfy our model given by Eqn. (\ref{eqn1}),
either for the direct hierarchy or the indirect hierarchy.
We first try the indirect hierarchy. In this case, the diagonal neutrino mass matrix, using the experimental mass difference squares, can be written as

\begin{equation} 
 \bold {M}_{\nu}^{diag}= \left(\begin{matrix}
\sqrt{m_3^2 + 0.002315} & 0 & 0 \\
0 & \sqrt{m_3^2 + 0.00239} & 0\\
0 & 0 & m_3
\end{matrix}
\right),  
\end{equation}
where we have used the definition of $\Delta m^2$ in the inverse hierarchy mode as referred in Table \ref{tab:expdata}.

Taking these experimental values in the best-fit($\pm\sigma$) region from Table \ref{tab:expdata},  for the PMNS mixing matrix, we get from Eqn (\ref{eqn2})
\begin{equation} 
 U_{\nu} = \left(\begin{matrix}
0.822 & 0.543 & 0.155 \exp(-i\delta) \\
-0.431-0.08 \exp(i\delta) & 0.652-0.05 \exp(i\delta) & 0.613\\
0.342-0.101 \exp(i\delta) & -0.517-0.07 \exp(i\delta) & 0.775
\end{matrix}
\right).  
\label{eqn3}
\end{equation}
We plug  these expressions  for $\bold {M}_{\nu}^{diag}$ and  $U_{\nu}$ in $ M_{\nu} = U_{\nu} \bold {M}_{\nu}^{diag} U_{\nu}^{\dagger}$ and 
demand that  the resulting mass matrix satisfy the form of our model predicted Eqn. (\ref{eqn1}). First, using
$M_{\mu \mu}$ = $M_{\tau \tau}$ as in Eqn. (\ref{eqn1}),  we obtain  the following  $2^{nd}$ order equation for  $\cos\delta$ 
\begin{eqnarray}
(12.78m_3^4 - 0.11m_3^2 - 0.0027)\cos^2\delta -(116.14m_3^4 - 6.98m_3^2 - 0.006)\cos\delta 
\nonumber\\ 
- 59.75m_3^4 + 3.56m_3^2  
 - 0.0034  = &0& ,
\label{equation}
\end{eqnarray}
where, we have used some approximations while simplifying the equation analytically, which would not affect our result, if it is done numerically. Further, Eqn. (\ref{equation}) is satisfied only for certain range of values of $m_3$  demanding that $ -1 < \cos\delta < 1 $.
For that range of $m_3$, now we demand  that 
$M_{e\mu}$ = $M_{e\tau}$ to be satisfied. This takes into account separately satisfying the equality of the real and imaginary parts of $M_{e\mu}$ and $M_{e\tau}$ elements. It is intriguing that a solution exists, and gives the values of 
$m_3 = 7.5 \times 10^{-2}$ eV and  $\delta = 110 \degree$.

Thus the prediction for the three neutrino masses and the CP violating phase in our model are,
\begin{eqnarray}
m_1 = \sqrt{m_3^2 + 0.002315} = 8.91 \times 10^{-2} eV ,\\\nonumber
m_2 = \sqrt{m_3^2 + 0.00239} = 8.95 \times 10^{-2} eV ,\\\nonumber
m_3 = 7.5 \times 10^{-2} eV ,\\\nonumber
\delta = 110 \degree .
\end{eqnarray}
with $\delta$ being close to the maximum CP violating phase.

As a double check of our calculation, we have calculated the neutrino mass matrix numerically using the above obtained values of $m_1, m_2 , m_3$ and $\delta$ as given by mass matrix  Eqn.(\ref{eqn0}).  The resulting numerical neutrino mass matrix we obtain is given by,  
\begin{equation} 
 M_{\nu} = \left(\begin{matrix}
0.088 & 0.00053 + 0.0011i & 0.00053 + 0.0017i \\
0.00053 - 0.0011i & 0.083 & -0.0065 \\
0.00053 - 0.0017i & -0.0065 & 0.081
\end{matrix}
\right).  
\end{equation}

We see that with this verification, the mass matrix predicted by our model in Eqn.(\ref{eqn1}), is well satisfied.


We note that we also investigated the normal hierarchy case for our model satisfying hermiticity and $\nu_{\mu} - \nu_{\tau}$ exchange symmetry. We found no solution for $\cos\delta$ for that case. Thus normal hierarchy for the neutrino masses  can not be accommodated in our model.

  Our model predicts the electron type neutrino mass to be rather large ($8.91 \times 10^{-2}$ eV), and the  CP violating parameter $\delta$ close to the maximal value ($\delta \simeq 110 \degree$). Let us now discuss briefly how our model can be tested in the proposed future experiments of electron type neutrino mass measurement directly and also for the leptonic CP violation. 
The measurement of the electron anti-neutrino mass from tritium $\beta$ decay in Troitsk $\nu$-mass experiment set a limit of $m_\nu < 2.2 eV$ 
 \cite {Aseev:2012zz}. 
New experimental approaches such as the MARE \cite{Schaeffer:2011zz} will perform  measurements of the neutrino mass in the sub-eV region. So with a little more improvement, it may be possible to reach our predicted value of $\sim 0.1$ eV. 

The magnitude of the CP violation effect depends directly on the magnitude of the well known Jarlskog Invariant \cite{Jarlskog:1985cw}, which is a function of the three mixing angles and CP violating phase $\delta$ in standard parametrization of the mixing matrix:
\begin{equation}
J_{CP} = 1/8 \cos\theta_{13} \sin2\theta_{12}\sin2\theta_{23}\sin2\theta_{13}\sin\delta
\end{equation}
Given the best fit values for the mixing angles in Table \ref{tab:expdata} and the value of CP violationg phase $\delta = 110\degree$ in our model, we find the value of Jarlskog Invariant,
\begin{equation}
J_{CP} = 0.033 , 
\end{equation}
which corresponds to large CP violating effects. The study of $\nu_\mu \rightarrow \nu_e$ and $\bar{\nu}_\mu \rightarrow  \bar{\nu}_e$ transitions using accelerator based beams is sensetive to the CP violating phenomena arising from the CP violating phase $\delta$. We are particularly interested in the Long Baseline Neutrino Experiment (LBNE) \cite{LBNE}, which with its baseline of $1300$ Km and neutrino energy $E_\nu$ between $1-6$ GeV would be able to unambiguously shed light both on the 
mass hierarchy and the CP phase simultaneously. Evidence of the CP violation in the neutrino sector requires the explicit observation of asymmetry between $P(\nu_\mu \rightarrow \nu_e)$ and $P(\bar{\nu}_\mu \rightarrow  \bar{\nu}_e)$, which is defined as the CP asymmetry $\mathcal{A}_{CP}$,
\begin{equation}
\mathcal{A}_{CP} = \frac{P(\nu_\mu \rightarrow \nu_e) - P(\bar{\nu}_\mu \rightarrow  \bar{\nu}_e)}{P(\nu_\mu \rightarrow \nu_e) + P(\bar{\nu}_\mu \rightarrow  \bar{\nu}_e)}
\end{equation}
In three-flavor model the asymmetry can be approximated to leading order in $\Delta m_{21}^2$ as, \cite{Marciano:2006uc}
\begin{equation}
\mathcal{A}_{CP} \sim \frac{\cos\theta_{23} \sin2\theta_{12} \sin\delta}{\sin\theta_{23}\sin\theta_{13}}(\frac{\Delta m_{21}^2 L}{4E_\nu}) + \text{matter effects} 
\end{equation}  
For our model, taking LBNE Baseline value L = $1300$ Km and $E_\nu = 1$ GeV, we get the value of $\mathcal{A}_{CP} = 0.162$ + matter effects.
With this relatively large values of $\mathcal{A}_{CP}$,
LBNE10 in first phase with values of 700 KW wide-band muon neutrino and muon anti-neutrino beams and $100$ kt.yrs will be sensitive to our predicted value of CP violating phase $\delta$ with 3-Sigma significance \cite{Kronfeld:2013uoa}.\\ 
Finally, we compare our model for the sum of the three neutrino masses against the cosmological observation.
The sum of neutrino masses $m_1 + m_2 + m_3 < (0.32 \pm 0.081)$ eV \cite {Battye:2013xqa} from (Planck + WMAP + CMB + BAO) for an active neutrino model with three degenerate neutrinos has become an important cosmological bound. For our model, we find $m_1 + m_2 + m_3 \simeq 0.25 $ eV, which is consistent with this bound.



\section{Summary and Conclusions}

In this work, we have presented a predictive model for Dirac neutrinos. 
The model has three assumptions: (i) lepton number conservation, (ii) hermiticity of the neutrino mass matrix, and (iii) $\nu_{\mu}$ - $\nu_{\tau}$ exchange symmetry. The resulting neutrino mass matrix is of Dirac type, and has five real parameters, (three real and one complex). We have shown that the data on neutrino mass differences squares, and three mixing angles are consistent with this model yielding a solution for the neutrino masses with inverted mass hierarchy (close the degenerate pattern). The values predicted by the model for the three  neutrino masses are  $8.91 \times 10^{-2}$ eV, $8.95 \times 10^{-2}$ eV and $7.50 \times 10^ {-2}$ eV. In addition, the model also predicts the CP violating phase $\delta$ to be $110 \degree$, thus predicting a rather large CP violation in the neutrino sector, and will be easily tested in the early runs of the LBNE. The mass of the electron type neutrino is also rather large, and has a good possibility for being accessible for measurement in the proposed tritium beta decay experiments. Neutrinos being Dirac, neutrinoless double beta decay is also forbidden in this model.  Thus, all of these predictions can be tested in the upcoming and future precision neutrino experiments.

\begin{acknowledgments}

This research was supported in part by  United States Department of Energy Grant Number DE-SC0010108.

\end{acknowledgments}


\end{document}